\documentclass{article}

\usepackage{arxiv}

\usepackage[utf8]{inputenc} 
\usepackage[T1]{fontenc}    
\usepackage{lmodern}        
\usepackage{hyperref}       
\usepackage{url}            
\usepackage{booktabs}       
\usepackage{amsfonts}       
\usepackage{nicefrac}       
\usepackage{microtype}      
\usepackage{lipsum}
\usepackage{graphicx}

\title{Linear Rescaling to Accurately Interpret Logarithms}

\author{
    Nick Huntington-Klein
    \thanks{nickchk.com}
   \\
    Department of Economics \\
    Seattle University \\
  Seattle University, 98122 \\
  \texttt{\href{mailto:nhuntington-klein@seattleu.edu}{\nolinkurl{nhuntington-klein@seattleu.edu}}} \\
  }

\newlength{\csllabelwidth}
\newlength{\cslhangindent}
  {}%
  {\par}
\newenvironment{CSLReferences}[2] 
 {
  \setlength{\parindent}{0pt}
  \ifodd #1 \everypar{\setlength{\hangindent}{\cslhangindent}}\ignorespaces\fi
  \ifnum #2 > 0
  \setlength{\parskip}{#2\baselineskip}
  \fi
 }%
 {}
\usepackage{calc} 

\usepackage{booktabs}
\usepackage{longtable}
\usepackage{array}
\usepackage{multirow}
\usepackage{wrapfig}
\usepackage{float}
\usepackage{colortbl}
\usepackage{pdflscape}
\usepackage{tabu}
\usepackage{threeparttable}
\usepackage{threeparttablex}
\usepackage[normalem]{ulem}
\usepackage{makecell}
\usepackage{xcolor}

\begin{document}
\maketitle

\def\tightlist{}

\begin{abstract}
The standard approximation of a natural logarithm in statistical analysis interprets a linear change of \(p\) in \(\ln(X)\) as a	\((1+p)\) proportional change in \(X\), which is only accurate for small values of \(p\). I suggest base-\((1+p)\) logarithms, where \(p\) is chosen ahead of time. A one-unit change in \(\log_{1+p}(X)\) is exactly equivalent to a \((1+p)\) proportional change in \(X\). This avoids an approximation applied too broadly, makes exact interpretation easier and less error-prone, improves approximation quality when approximations are used, makes the change of interest a one-log-unit change like other regression variables, and reduces error from the use of \(\log(1+X)\).
\end{abstract}


\hypertarget{the-traditional-interpretation-of-logarithms}{%
\section{The Traditional Interpretation of
Logarithms}\label{the-traditional-interpretation-of-logarithms}}

It is common practice in many statistical applications, especially in
regression analysis, to transform variables using the natural logarithm
\(\ln(X)\). This can be done for statistical reasons, for example to fit an apparent
functional form in the data or to reduce skew and the impact of positive
outliers in the variable \(X\). The logarithm transformation is also
used for theoretical reasons, when theory dictates the model relates to
proportional changes in \(X\) rather than linear changes.

The standard interpretation of a log-transformed variable in a
regression is that a linear increase of \(p\) in \(\ln(X)\) is
equivalent to a \(p\times 100\%\) increase in \(X\). This is not
literally true. A linear increase of \(p\) in \(\ln(X)\) is equivalent
to an \((e^p - 1)\times 100\%\) increase in \(X\). The standard
interpretation relies on the approximation \(e^p \approx 1+p\), or
equivalently \(p \approx \ln(1+p)\), which is fairly accurate for small
values of \(p\).

\begin{equation}
	\ln(X) + p \approx \ln(X) + \ln(1+p)  = \ln(X(1+p))
\end{equation}

In this paper, I provide an very simple alternative approach to using
and interpreting logarithms in the context of regression analysis, which
solves three major problems with the standard approach.

The first major problem with the standard approach is that the
approximation loses quality relatively quickly as \(p\) grows. The error
in approximation is equal to

\[ 1+p - e^p  \]

which is always negative, and grows more negative with \(p\), such that
this approximation always understates the proportional increase in \(X\)
equivalent to a given linear increase in \(\ln(X)\). If \(\ln(X)\) is a
treatment variable, the approximation will always overstate its effect.

The quality of approximation is, subjectively, acceptable for small
values of \(p\), but the error becomes large within ranges of interest.
For linear increases in \(\ln(X)\) of .1, .2, or .3, respectively,
interpretations of these changes as 10\%, 20\%, or 30\% increases in
\(X\) would understate the actual change in \(X\) by about .5, 2.1, and
5 percentage points, respectively.\footnote{The approximation error expression also
	demonstrates that \(e\) does not hold any special property in regards to
	reducing approximation error, and is actually a poor choice of logarithmic base if
	the traditional approximation is to be used. Many other bases, like
	2.6, produce nearly identical errors for small \(p\) and then dominate
	\(e\) afterwards. Base
	2.35 is attractive in other ways: errors for base 2.35 are no larger in
	absolute value than .014 all the way up to \(p = .43\), and
	considerably improve on base \(e\) above that, although unfortunately
	performance relative to base-\(e\) is worst around \(p=.1\). If a
	researcher insists on the traditional approximation, I at least
	recommend the use of base-2.6 logarithm, which is a clear improvement
	on \(e\), or perhaps base-2.35 for something less sensitive to \(p\).}

This leads naturally to the second problem with the standard
interpretation, which is sociological in nature. The fact that the
base-\(e\) approximation breaks down quickly as \(p\) increases does not
appear to be universally known, nor is there a standard maximum \(p\)
for which the approximation is considered acceptable. It is not uncommon
to see papers describing 10\% or 20\% changes in a log-transformed
variable, and it is doubtful that these authors would willingly inject
biases of .5 or 2.1 percentage points into their analysis for no reason.

Calculations that give exact interpretations using \(e^p\) or
\(\ln(1+p)\) are available but are not universally applied even for
larger percentage changes. Speculatively, this may be because the author
assumes that approximation error is too small to bother, because the
additional calculation could confuse a reader, or because in some fields it is not expected. It may even be the case that the
researcher is not aware that the traditional interpretation \emph{is} an
approximation. It is common in published studies and in econometric
teaching materials to see the traditional
\(\ln(X)+p \approx \ln(X(1+p))\) approximation discussed without
reference to its approximate nature, implying by omission that it is an
equality. At the undergraduate level, see for example
Bailey (2017) Section 7.2. At the graduate level, see
Greene (2008) Section 4.7, specifically Example 4.3.

The third problem with the standard interpretation, when applied to
variables on the right-hand side of a regression, is that it requires
nonstandard interpretation of regression coefficients. Nearly all
regression coefficients are understood in terms of one-unit changes in
the associated predictor. Log-transformed variables are an exception to
this. A one-unit change in \(\ln(X)\) describes what would in most cases
be an unrealistically large increase in \(X\), and also would produce a
large 71.8 percentage point error if the traditional approximation were
applied.

All three of these problems can be solved by simply changing the base of
the logarithm.\footnote{After a literature search, I was unable to find
	previous studies making this same recommendation. However, given the
	long history of logarithms in regression, it seems unlikely that
	nobody has thought of the insight in this paper before. So I will not
	claim that this method is novel, but just that it is currently not
	widely known or applied.} Selecting a percentage increase
\(p\times 100\%\) ahead of time and using \(\log_{1+p}(X)\) in place of
\(\ln(X)\) means that a one-unit change in \(\log_{1+p}(X)\) is exactly
equivalent to a \(p\times 100\%\) increase in \(X\). There is no need
for approximation, and the exact interpretation can be written directly
into a regression table, solving the first problem and much of the
second. The use of base \(1+p\) can be restated as \(\ln(X)/\ln(1+p)\),
framing the method in the easily-understood terms of linearly rescaling
the variable by the constant \(1/\ln(1+p)\). The third problem is also
solved because the relevant increase in \(\log_{1+p}(X)\) is 1, in line
with other variables in the regression.

Based on these results, I recommend the use of alternate logarithmic
bases, or the ``linear rescaling'' approach, when using logarithms in
statistical analysis, especially in regression. The benefits are most
apparent when applied to variables on the right-hand side of a
regression (the predictors/independent variables), but there are also
benefits on the left-hand side (the outcome). Additionally, the use of
linear rescaling helps ease some problems related to the use of
\(\ln(1+X)\) with \(X\) variables that contain values of zero.

\hypertarget{linearly-rescaling-logarithms}{%
\section{Linearly Rescaling
Logarithms}\label{linearly-rescaling-logarithms}}

\hypertarget{exact-interpretation-of-logarithms}{%
\subsection{Exact Interpretation of
Logarithms}\label{exact-interpretation-of-logarithms}}

\label{sec:exact}

For any base \(b\), a linear increase of \(p\) in a logarithm
\(\log_b(X)\) is equivalent to a proportional change in \(X\) of
\(b^p\), or a percentage increase of \((b^p-1)\times 100\%\).
Researchers could report exact interpretation of linear logarithmic
increases using this formula. However, this practice is far from
universal.

Another approach to exact interpretation of linear logarithmic increases
is to take advantage of the following feature of \(b^p-1\):

\[ (1+p)^1 - 1 = p \]

That is, a linear increase of \(1\) in \(\log_{1+p}(X)\) is exactly
equivalent to a percentage increase of \(p\times100\%\) (or a
proportional increase of \(1+p\)) in \(X\).

This means that a researcher can select ahead of time the percentage
increase in \(X\) that they are interested in, for example 10\%
(although any other percentage would work as well), and then use
\(\log_{1.1}(X)\) in their analysis instead of \(\ln(X)\). Then, a
one-unit increase in \(\log_{1.1}(X)\) can be exactly interpreted as a
10\% increase in \(X\).

Further, because of the change of base formula,
\(\log_{1.1}(X)\) can be calculated as \(\ln(X)/\ln(1.1)\).
\(\ln(1.1)\) is a constant, and so the researcher can achieve exact
interpretation of the change by scaling the variable they're already
using (\(\ln(X)\)) by a constant. Researchers using any estimation
method should already be aware of the implications of scaling by a
constant in that method, so the linear rescaling approach to changing
the base should be understandable by both researchers and readers of
research. The use of the change-of-base formula also allows another
clear demonstration of what the choice of logarithm base does for
interpretations of proportional change:

\[ \frac{\ln(X(1+p))}{\ln(1+p)} = \frac{\ln(X) + \ln(1+p)}{\ln(1+p)} = \frac{\ln(X)}{\ln(1+p)}+1 \]

There are several benefits to linear rescaling:

\begin{itemize}
	\item
	It produces exact interpretations of linear increases in logarithms.
	\item
	It is, arguably, conceptually more simple than using \(b^p-1\) or
	\(log_b(1+p)\) to adjust the result after estimation, and avoids the introduction of another calculation where error may occur.
	\item
	The change of interest is one log unit, which is how most other
	variables are understood.
	\item
	The exact interpretation can be written directly onto a regression
	table rather than relying on supplemental calculations, as will be
	shown in the following sections.
\end{itemize}

The main downside of this approach is that it requires the choice of the
percentage increase beforehand. In practice, this is unlikely to be a
major hurdle, as researchers often report only a single percentage
increase value anyway, typically a preselected value like 10\%, based on
what a reasonable observable change in \(X\) would be.

Additionally, if selecting a percentage increase beforehand is not
realistic, or if multiple percentage increases are desired, exact
interpretation is still available, as in the traditional method, using
\((1+b)^{p}-1\). Linear rescaling also improves the process of adjusting
the estimate, at least on the right-hand side of a regression (see
Section \ref{sec:rhs}).

Further, if an approximation is used instead of an exact interpretation,
linear rescaling often produces better approximations than the
traditional approximation. For example, if \(\log_{1.1}(X)\) has been
used and the researcher wants to approximate a 20\% increase in \(X\)
using a two-unit increase in \(\log_{1.1}(X)\), they will actually see
the effects of a \(1.1^2=1.21\), or 21\%, increase, rather than 20\%.
This is an error of one percentage point, compared to the traditional
approach, which produces an error of 2.1 percentage points for a .2
increase.

\begin{figure}
	\centering
	\includegraphics[width=.75\textwidth]{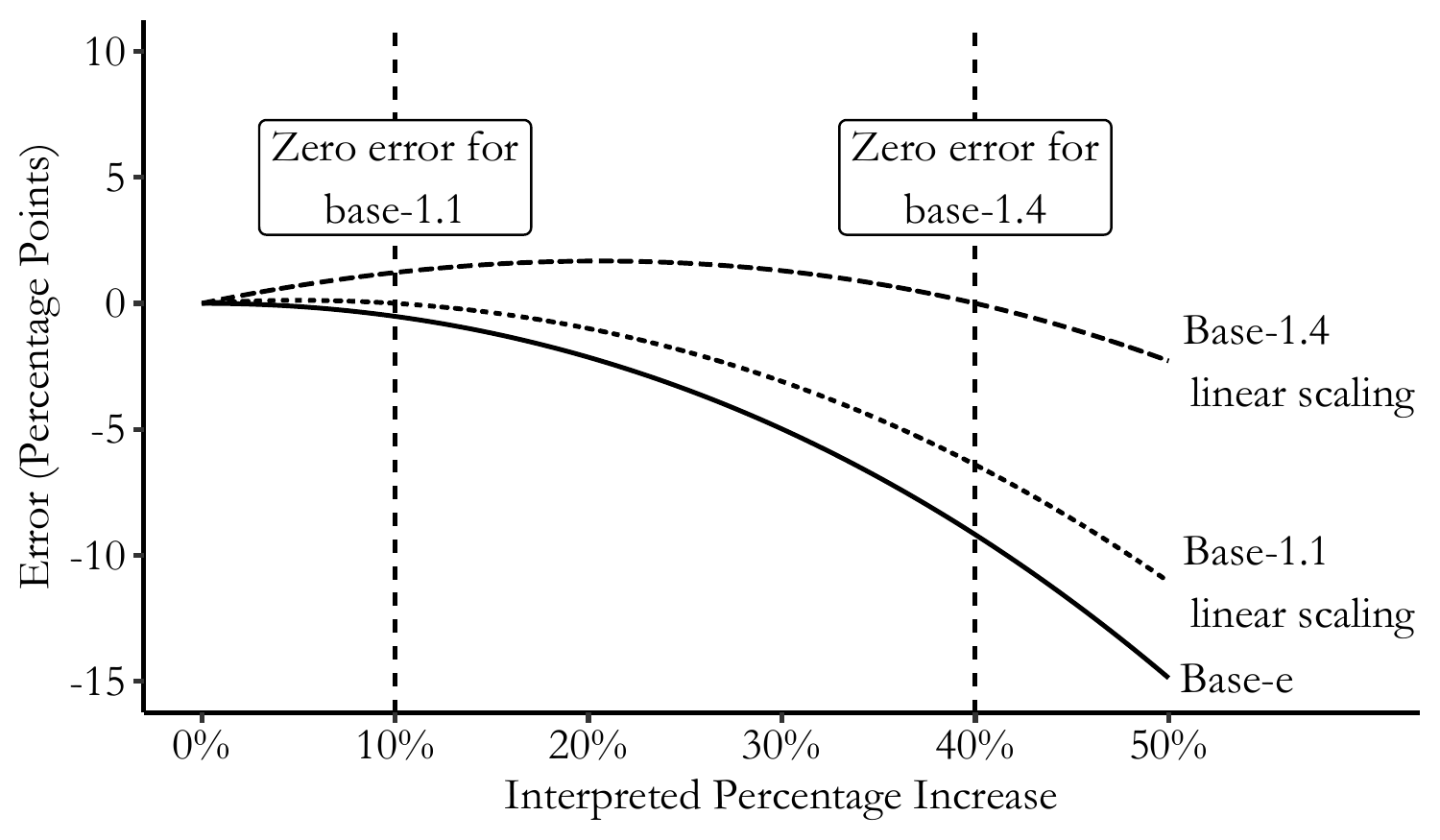}
	\caption{\label{fig:approxquality} Approximation Error with Traditional
		and Linear Rescaling Methods}
\end{figure}

Figure \ref{fig:approxquality} examines the error in approximating
different percentage increases with a traditional approximation, using a
base-\(e\) logarithm where a \(p\)-unit increase in \(\ln(X)\) is taken
to be a \(p\times 100\%\) increase in \(X\). I contrast the traditional
approximations with approximations from the linear rescaling approach
using two different bases: a base-\(1.1\) logarithm, where a \(p\)-unit
increase in \(\log_{1.1}(X)\) is taken to be a \(p\times 10\%\) increase
in \(X\), and a base-\(1.4\) logarithm, where a \(p\)-unit increase in
\(\log_{1.4}(X)\) is taken to be a \(p\times 40\%\) increase in \(X\).

The traditional base-\(e\) method slightly outperforms the base-1.1
linear rescaling method, by a miniscule degree, up to a linear change of
.048 (although the linear rescaling method could outperform the
traditional method for any linear change by selecting a different base
than the ones shown in the graph). After .048, approximation with
base-1.1 linear rescaling dominates the traditional approximation,
especially near \(p=.1\). Both the traditional and base-1.1
approximations considerably outperform base-1.4 for small \(p\), but
this is to be expected - base-1.4 is to be used when the change of
interest is 40\%. In the region of \(p=.4\), the base-1.4 logarithm
considerably outperforms the other two, and approximation errors with
base-1.4 are relatively small for the entire graphed range up to
\(p=.5\). While the linear rescaling method allows for easy access to
exact interpretation, even in cases where approximation is used, the
linear rescaling approximation error will be smaller than the
approximation error for the traditional method as long as the linear
rescaling base is near enough to the proportional change of interest.

\hypertarget{linear-rescaling-on-the-right-hand-side}{%
\subsection{Linear Rescaling on the Right Hand
Side}\label{linear-rescaling-on-the-right-hand-side}}\label{sec:rhs}

The benefits of linear rescaling are clearest when applied to a
logarithmic transformation on the right-hand side of a regression.
Considering the model:

\[ Y = \beta_0 + \beta_1\ln(X) + \varepsilon \]

the interpretation of \(\hat{\beta}_1\) is often given in a format
similar to ``a 10\% increase in \(X\) is associated with a
\(.1\times\hat{\beta}_1\) increase in \(Y\).'' Or for an exact
interpretation, ``a 10\% increase in \(X\) is associated with a
\(\ln(1.1)\times\hat{\beta}_1 = .0953\times\hat{\beta}_1\) increase in
\(Y\).''

Under linear rescaling for a 10\% increase in \(X\), instead the model
is:

\[ Y = \beta_0 + \beta_1\frac{\ln(X)}{\ln(1.1)} + \varepsilon \]

in which the interpretation of \(\hat{\beta}_1\) is the simpler ``a 10\%
increase in \(X\) is associated with a \(\hat{\beta}_1\) increase in
\(Y\).''

The interpretation is simple enough that it can be written directly into
a regression table, as in Table \ref{tab:regtable} Column 1, rather than relying on additional in-text
calculations. The row heading in the table itself ``\(X\) (10\%
Increase)'' is able to convey that a 10\% change in \(X\) is associated
with a \(\hat{\beta}_1\) change in \(Y\), and the table note provides
more detail. The label ``\(\log_{1.1}(X)\) (10\% increase)'' may be
preferred.

If the researcher is interested in multiple percentage changes,
adjustment to get exact interpretation for each percentage change is
straightforward under linear rescaling, because of the change-of-base
formula. If \(\log_{1.1}(X)\) has been used, but the researcher wants an
exact interpretation for the linear change equivalent to a 20\% increase
in \(X\), the researcher can multiply the coefficient \(\hat{\beta}_1\)
by \(\ln(1.2)/\ln(1.1)\). This will work for any regression method where
scaling a predictor by \(c\) has the result of scaling its coefficient
by \(1/c\).

\begin{table}
	
	\caption{\label{tab:regtable}Example Regression Table with Linear Rescaling}
	\centering
	\begin{tabular}[t]{lccc}
		\toprule
		& Y & Y (10\% Increase) & Y (10\% Increase) \\
		\midrule
		X (10\% Increase) & 0.194*** &  & 0.460***\\
		& (0.003) &  & (0.007)\\
		X &  & 2.001*** & \\
		&  & (0.038) & \\
		\midrule
		Num.Obs. & 1000 & 1000 & 1000\\
		\bottomrule
		\multicolumn{4}{l}{\rule{0pt}{1em}* p $<$ 0.1, ** p $<$ 0.05, *** p $<$ 0.01}\\
		\multicolumn{4}{l}{\textsuperscript{} Variables marked with 10\% increase use a base-1.1 logarithm}\\
		\multicolumn{4}{l}{transformation. Data is simulated.}\\
	\end{tabular}
\end{table}

\hypertarget{linear-rescaling-on-the-left-hand-side}{%
\subsection{Linear Rescaling on the Left Hand
Side}\label{linear-rescaling-on-the-left-hand-side}}

\label{sec:lhs}

The benefits of linear rescaling are less clear on the left-hand side of
the regression, since the proportional change of interest cannot be
exactly chosen ahead of time. Still, there are benefits.

Consider Table \ref{tab:regtable} Column 2, which uses the model

\[ \frac{\ln(Y)}{\ln(1.1)} = \beta_0 + \beta_1X + \varepsilon \] The
regression estimate is \(\hat{\beta}_1 = 2.001\). By itself, this does
not easily lead to exact interpretation of the coefficient.

At this point, the researcher can approximate the effect of \(2.001\) as
a \(2.001\times 10\% \approx 20\%\) increase. As in Section
\ref{sec:exact}, this will lead to less approximation error than in the
traditional method provided that \(\hat{\beta}_1\) is not too far from
\(1\).

There is also the option to provide an exact interpretation, where a
one-unit increase in \(X\) is associated with a \(b^{\hat{\beta}_1}\)
proporional change in \(Y\). This is not much different from the process
for getting exact interpretation using the traditional approach,
although it may be somewhat easier to understand if \(p\) is a more
natural object to think about than \(e\).

A third option is to rerun the model with a different logarithmic base
such that the \(\hat{\beta}_1\) will be near \(1\). Then,
\(\hat{\beta}_1\) can be very accurately interpreted as a proportional
increase of \(b\) in \(Y\). However, this is both laborious and would
result in the coefficient of interest being oddly located in the
logarithmic base.

One note about left-hand side use is that the traditional approximation
is known to perform particularly poorly, and exact interpretation is
especially important, when the logarithm is on the left-hand side and a
predictor of interest is binary (Halvorsen \& Palmquist, 1980). In
theoretical terms this is because the derivative does not exist. In
practical terms this is because, if the coefficient on the binary
variable is large, the researcher cannot naturally select a linear
change small enough for the approximation to perform well. Easy access
to exact interpretation, and improved approximation when used, are
especially important in this case.

\hypertarget{linear-rescaling-on-both-sides}{%
\subsection{Linear Rescaling on Both
Sides}\label{linear-rescaling-on-both-sides}}

\label{sec:bothsides}

In the case of the log-log model

\[ \ln(Y) = \beta_0 + \beta_1\ln(X) + \varepsilon \]

linear rescaling on both the left and right-hand sides using the same
bases will have no effect on \(\hat{\beta}_1\) or on its interpretation,
and offers no major improvement over the traditional method, unless
there is a reason to want different bases on the left and right-hand
sides, or if there is interest in interpreting the coefficients on
control variables with a linearly-rescaled left-hand side.

There are some minor expositional benefits. Linear rescaling could be
used here for consistency with other models that are not log-log.
Rescaling can also make clear to an audience unfamiliar with log-log
models how they can be interpreted. For example, it's not uncommon in a
log-log model to still report a result like ``a 10\% increase in \(X\)
is associated with a \(\hat{\beta}_1\times .1\)\% change in \(Y\).''
Herz \& Mejer (2016) is just one example of this. If the author wants the
reader to think in terms of a percentage increase of a particular size
in this way, linear rescaling can make that interpretation explicit on
the regression table, as in Column 3 of Table \ref{tab:regtable}.

\hypertarget{linear-rescaling-and-zeroes}{%
\subsection{Linear Rescaling and
Zeroes}\label{linear-rescaling-and-zeroes}}

Researchers often want to apply a logarithmic transformation to a
variable that can take values of zero. There are two common
approaches to this: \(\ln(1+X)\), and the asymptotic hyperbolic sine
transformation \(asinh(X) = \ln(X+\sqrt{X^2+1})\). Exact calculations
for elasticity interpretations using the \(asinh(X)\) transformation are
described in Bellemare \& Wichman (2020). Both \(\ln(1+X)\) and
\(asinh(X)\) reduce skew and accept values of zero. However, if the researcher
wishes to maintain a proportional-change interpretation (at values other
than zero), there are problems with any sort of ad-hoc
transformation like this, including a sensitivity to the scale of \(X\)
and the fact that the zero-censored variable is treated as uncensored.
For a left-hand side variable, poisson regression or a censoring model
are likely to be superior to an ad-hoc transformation. However, the use
of an ad-hoc transformation is still a concern on the right-hand side or
for researchers who want to use standard linear regression for other
reasons.

In this section, I will assume that the researcher's goal is interpret a
linear change in \(\ln(1+X)\) in terms of a proportional change in \(X\)
(rather than a proportional change in \(1+X\)). In this case, exact
interpretation is particularly important, whether performed using linear
rescaling or \(e^p\), but there are several details that still separate
the two approaches.

Consider a one-unit increase in \(\ln(1+X)/\ln(1+p)\). This is
equivalent to a proportional change of \(1+p\) in \(1+X\), or an
absolute increase of \(p(1+X)\). If \(X\) increases by \(p(1+X)\), what
proportional increase is that equivalent to?

\begin{equation}
	\label{eq:linerror}
	X + p(1+X) = X(1+p + \frac{p}{X})
\end{equation}

A one-unit linear increase in \(\ln(1+X)/\ln(1+p)\) interpreted as a
\(1+p\) proportional change in \(X\) will get the proportional change
wrong by \(\frac{p}{X}\).

Similarly, a linear increase of \(p\) in \(\ln(1+X)\) is a proportional
increase of \(e^p\) in \(1+p\). As above,

\begin{equation} 
	\label{eq:traderror}
	X+(e^p-1)(1+X) = X(e^p + \frac{e^p-1}{X})
\end{equation}

A linear increase of \(p\) in \(\ln(1+X)\) interpreted as a proportional
increase of \(e^p\) in \(X\) will get the proportion wrong by
\(\frac{e^p-1}{X}\).

For a given \(p\), since \(p \leq e^p-1 \ \forall \ p \geq 0\), the
linear rescaling approach will always outperform the traditional
approach, and by a greater margin as \(p\) increases. However, this is
due to the fact that for a given \(p\), the traditional method describes
a proportional change of \(e^p \geq 1+p\). For a given
\emph{proportional change}, for example comparing a linear increase of
\(1\) under linear rescaling to a linear increase of \(\ln(1+p)\) in the
traditional method, both methods perform identically.

There are still several points to recommend linear rescaling here.
First, linear rescaling is an improvement if researchers using the
traditional method first select a \(p\) of interest and then calculate
\(e^p\), rather than selecting \(e^p\) directly (since, as above, \(p \leq e^p-1 \ \forall \ p \geq 0\)). Second, a bias of
\(p/X\) may be easier to reason about than \((e^p-1)/X\).

The comparison so far assumes that the researcher using the traditional
approach uses the exact interpretation, where a linear increase of \(p\)
is a proportional increase of \(e^p\). If they instead interpret a
linear increase of \(p\) as a proportional change of \(1+p\) in \(X\)
using the \(e^p \approx 1+p\) approximation, the error will be

\[ e^p - (1+p) + \frac{e^p-1}{X}  \]

There are two problems here. First, the fact that linear rescaling and
the traditional method perform identically for a given proportional
increase doesn't matter, as by using the approximation the researcher
has chosen to fix \(p\), under which linear rescaling outperforms the
traditional method. Second, the use of the approximation adds the
traditional approximation error \(e^p-(1+p)\) on top of
\(\frac{e^p-1}{X}\), further increasing error relative to the linear
rescaling method, and making the error grow even faster in \(p\).

For either the linear rescaling or traditional approaches, the
recommendation from Bellemare \& Wichman (2020) to scale \(X\)
upwards in reference to \(asinh(X)\), elaborated upon in
Aihounton \& Henningsen (2019) to determine optimal scaling values, is also
implied by these results for \(\ln(1+X)\), as the interpretation error
declines proportionally with larger absolute values of \(X\).

\hypertarget{exact-interpretation-with-zeroes}{\subsection{Exact interpretation with
		zeroes}\label{exact-interpretation-with-zeroes}}

As an aside (since it does not relate to linear rescaling in
particular), a researcher using either the linear rescaling or
traditional method could use one of the error formulas in Equations
\ref{eq:linerror} or \ref{eq:traderror} to adjust their
proportional-change interpretation, or decide whether the error is small
enough to ignore, for a given \(p\) and \(X\).

This may be especially useful in the calculation of elasticities, since
it allows proportional changes in both \(Y\) and \(X\) to be recovered
from proportional changes in \(1+Y\) and \(1+X\).

For example, in the log-log model

\[\frac{\ln(1+Y)}{\ln(1+p_Y)} = \beta_0 + \beta_1\frac{\ln(1+X)}{\ln(1+p_X)} + \varepsilon  \]

a \(1+p_X\) proportional change in \(1+X\) is associated with a
\((1+p_Y)^{\hat{\beta}_1}\) proportional change in \(1+Y\). For the
specific values \(X = X_0\) and \(Y = Y_0\), this means that a
\(1 + p_X + \frac{p_X}{X_0}\) proportional change in \(X\) is associated
with a
\((1+p_Y)^{\hat{\beta}_1} + \frac{(1+p_Y)^{\hat{\beta}_1}-1}{Y_0}\)
proportional change in \(Y\). If desired, \(p_X\) and \(X_0\) can be
selected ahead of time so that \(1 + p_X + \frac{p_X}{X_0}\) is a round
number. Similar calculations follow for the \(\ln(1+X)\)-linear and
linear-\(\ln(1+X)\) cases. The only task remaining at that point is
calculation of standard errors for this nonlinear function of
\(\hat{\beta}_1\). The delta method is one acceptable approach, at least
in large samples.\footnote{Keep in mind that this adjustment does not
	account for zero-censoring of the variables, which is still present
	and may still harm performance for logarithms on the left-hand side
	relative to using a model that properly accounts for censoring.}

\hypertarget{example-applications}{\section{Example Applications}
	\label{sec:examples}}

In this section I refer to several published studies that use natural
logarithm transformations in regression analysis, and discuss how those
papers might have been different using linear rescaling.

The first study I look at, which is from economics, is
Eren, Onda, \& Unel (2019). This study looks at the impact of foreign
direct investment (FDI) on entrepreneurship in the United States,
covering business-creation, business-destruction, and self-employment
rates as outcome variables. The authors use a natural logarithmic
transformation of FDI, and the headline result specified in the abstract
is ``A 10\% increase in FDI decreases the average monthly rate of
business creation and destruction by roughly 4 and 2.5\% (relative to
the sample mean), respectively,'' where these figures refer
to results in states without Right-to-Work laws. Despite the percentage
interpretation given for the effects, business creation and destruction
are not log-transformed, only FDI is. The linear effect of log FDI is
reported as a percentage of the sample means of business creation and
destruction.

This study makes use of the traditional approximation. The result for
business creation comes from a regression of business creation rates on
\(\ln(FDI)\) (lagged two years), where the coefficient on \(\ln(FDI)\)
is -1.083. They interpret this as a 10\% increase in FDI reducing the
business creation rate by .1083 (or perhaps they round the effect to .11
before proceeding, this is not clear), which is roughly 4\% of the mean
(\(.1083/2.889 = .0375\) or \(.11/2.889=.038\)). Under linear rescaling
with a base of \(1.1\), the coefficient on \(\log_{1.1}(FDI)\) would be
\(.1032\), which indicates an effect roughly 3.5\% of the mean
(\(.1032/2.889=.0357\)), keeping in mind that the abstract reports its
other effect to the half-percent level of precision. In this case, the
use of the approximation made the effect seem larger than it was. The
correction is not enormous, but still there is no particular reason the
correct result could not have been in the original study. Traditional
exact interpretation or linear rescaling would have avoided this
problem. However, linear rescaling in this case offers the additional
benefit over traditional exact approximation that, if the dependent
variable had been divided by its mean, the coefficient on
\(\log_{1.1}(FDI)\) would have been the value of interest \(-.0357\)
directly, and no further calculation would have been necessary. This
ability to put the value of interest on the table also applies to the
2.5\% result, although in this case the substantive reported conclusion
does not change at the half-percent level of precision (the effect to
two decimal places drops from to 2.55\% to 2.43\%).

The second study is from public health. Kim \& Leigh (2020) look
at the effect of wages on obesity rates, finding that low wages increase
body mass and the prevalence of obesity. This study uses traditional
exact interpretation, so linear rescaling cannot change its results, but
it may change how they are presented. Their instrumental variables
estimate with BMI as a dependent variable reads ``the coefficient on
ln-wage was statistically significant (\(p < 0.01\)) and its value was
\(-3.3\). Standard errors appear in parentheses. This coefficient
suggests that a 10\% increase in wages is associated with 0.32 decline
in BMI.'' The .32 appears to be a slight miscalculation and can be
derived from \(\ln(1.1)\times(-3.3) = -.3145\), which should round to
\(-.31\), not \(-.32\). Under linear rescaling with a base of \(1.1\),
this additional calculation would not need to appear in the text, and
the coefficient on \(\log_{1.1}(Wage)\) would be rounded to \(-.3145\)
or \(-.315\), which would also avoid the slight miscalculation. The
authors also report the effect of a 100\% increase in wages as a 2.29
decline in BMI (\(\ln(2)\times(-3.3) = -2.29\)). To achieve this with
linear rescaling, the authors either could rerun analysis using a log
base of 2, or, more realistically, retain the calculation in the text
for this additional result, adjusting the 10\% estimate by
\(-.3145\times\ln(2)/\ln(1.1) = -2.29\). This simplified presentation
for the BMI results could be similarly applied in analysis of the
obesity rate dependent variable.

The final study I will mention is an example of how the complex
calculations necessary to produce exact interpretations of logarithms
using the traditional method can lead to error. Lin, Teymourian, \& Tursini (2018)
look at the effect of the natural logarithm of sugar and processed food
imports on the prevalence of obesity and overweight. In one model the
coefficient on logged imports is .085, which is interpreted as ``10\%
increase in import is associated with approximately 0.004 increase in
average BMI.'' However, the effect should instead be of a .0081 increase
in BMI using exact interpretation, or .0085 using the traditional
approximation, an effect more than twice as large. Similar errors are
made in another table, where a coefficient of .004 is again interpreted
as having about half the appropriate effect size for a 10\% increase.
They also provide an interpretation of a 50\% increase, which is about
40\% the appropriate size. If the authors had used linear rescaling,
they would not have needed to produce these calculations, and there
would have been no potential for this error to occur. Instead, the
effect size of interest would have automatically been produced by the
statistics software.

\hypertarget{conclusion}{%
\section{Conclusion}\label{conclusion}}

Despite their wide use across many fields, both in the context of
regression and in other statistical applications, logarithms are
frequently misinterpreted to greater or lesser degrees. This is partly
due to the standard interpretation of logarithms, which relies on an
approximation that produces non-negligible errors outside of a fairly
narrow range of percentage increases. It is possible to skip the
approximation and instead accurately interpret a linear increase \(p\)
in \(\ln(X)\) as a proportional increase of \(e^p\) in \(X\). However,
this practice is not universal, and carries a chance of error.

The interpretation of logarithms can be improved by changing the base of
the logarithm to \(1+p\), where \(1+p\) is a proportional change of
interest. The change of base can be achieved by linearly rescaling
\(\ln(X)\) to \(\ln(X)/\ln(1+p)\). This rescaling offers a way of
producing exact interpretations of logarithms, or in some cases
approximations with less error than the traditional approximation.

Linear rescaling can be easier for a reader to understand on a
regression table. The one-unit log increase that accompanies a given
percentage increase can be understood in the same way as changes in
untransformed variables. In order to interpret the coefficient the
reader does not need to search for additional calculations in the text,
nor does the author need to perform and provide them.

Crucially, rescaling is as easy to perform and explain as the
traditional approximation, and does not in most cases require the
additional post-analysis calculations that usually go along with exact
interpretations under the traditional method. All three of the studies
covered in Section \ref{sec:examples} contained either errors or
misrepresentations of their results because of these post-analysis
calculations. I did not select these studies because I knew they had
problems or anticipated any; it just so happens that the first three
studies I selected as good candidates for demonstrating the method all
had problems that could have been avoided with linear rescaling.

Ease of use for the researcher is important, because it may help
sidestep some of the reasons why researchers do not already report exact
results, and may help avoid calculation errors with traditional exact
interpretation. Researchers should in general be producing exact
interpretations of logarithms and be aware of the extent of error in the
traditional approximation. Because it is as easy as the traditional
approximation, linear rescaling is an attractive way of providing exact
results.

\hypertarget{references}{%
\section*{References}\label{references}}
\addcontentsline{toc}{section}{References}

\hypertarget{refs}{}
\begin{CSLReferences}{1}{0}
\leavevmode\vadjust pre{\hypertarget{ref-aihounton2019units}{}}%
Aihounton, Ghislain BD, and Arne Henningsen. 2019. {``Units of
Measurement and the Inverse Hyperbolic Sine Transformation.''} IFRO
Working Paper.

\leavevmode\vadjust pre{\hypertarget{ref-bailey2017real}{}}%
Bailey, Michael A. 2017. \emph{Real Econometrics: The Right Tools to
Answer Important Questions}. Oxford University Press.

\leavevmode\vadjust pre{\hypertarget{ref-bellemare2020elasticities}{}}%
Bellemare, Marc F, and Casey J Wichman. 2020. {``Elasticities and the
Inverse Hyperbolic Sine Transformation.''} \emph{Oxford Bulletin of
Economics and Statistics} 82 (1): 50--61.

\leavevmode\vadjust pre{\hypertarget{ref-eren2019}{}}%
Eren, Ozkan, Onda, Masayuki, and Bulent Unel. 2019. {``Effects of FDI on Entrepreneurship: Evidence from Right-to-work and Non-right-to-work States.''} \emph{Labour Economics} 58: 98--109.

\leavevmode\vadjust pre{\hypertarget{ref-greene2003econometric}{}}%
Greene, William H. 2008. \emph{Econometric Analysis}. Pearson Education.

\leavevmode\vadjust pre{\hypertarget{ref-halvorsen1980interpretation}{}}%
Halvorsen, Robert, and Raymond Palmquist. 1980. {``The Interpretation of
Dummy Variables in Semilogarithmic Equations.''} \emph{American Economic
Review} 70 (3): 474--75.

\leavevmode\vadjust pre{\hypertarget{ref-herz2016}{}}%
Herz, Benedikt, and Malwina Mejer. 2016. {``On the Fee Elasticity of the Demand for Trademarks in Europe.''} \emph{Oxford Economic Papers} 68 (4): 1039--1061.

\leavevmode\vadjust pre{\hypertarget{ref-kim2010estimating}{}}%
Kim, DaeHwan, and John Paul Leigh. 2010. {``Estimating the Effects of Wages on Obesity.''} \emph{Journal of Occupational and Environmental Medicine} 52 (5): 495--500.

\leavevmode\vadjust pre{\hypertarget{ref-lin2018effect}{}}%
Lin, Tracy Kuo, Teymourian, Yasmin, and Maitri Shila Tursini. 2018. {``The Effect of Sugar and Processed Food Imports on the Prevalence of Overweight and Obesity in 172 Countries.''} \emph{Globalization and health} 14 (1): 1--14.

\end{CSLReferences}

\end{document}